\documentclass[lettersize,journal]{IEEEtran}
\usepackage{amsmath,amsfonts}
\usepackage{algorithmic}
\usepackage{algorithm}
\usepackage{array}
\usepackage{subcaption} 

\usepackage{textcomp}
\usepackage{stfloats}
\usepackage{url}
\usepackage{makeidx}
\usepackage{verbatim}
\usepackage{tabularx}
\usepackage{xcolor} 

\definecolor{darkblue}{rgb}{0,0,0.5}
\usepackage{graphicx}
\usepackage{cite}
\usepackage{nomencl}
\makenomenclature

\usepackage{blkarray}
    
    \usepackage{cuted} 
    \usepackage{glossaries}
    \makeglossaries

\usepackage{xcolor}
\usepackage{tikz}
\usetikzlibrary{shapes.geometric, arrows}

\tikzstyle{startstop} = [rectangle, rounded corners, 
minimum width=3cm, 
minimum height=1cm,
text centered, 
text width=5cm, 
draw=black, 
fill=white!30]

\tikzstyle{stop} = [rectangle, rounded corners, 
minimum width=3cm, 
minimum height=1cm,
text centered, 
draw=black, 
fill=white!30]

\tikzstyle{io} = [trapezium, 
trapezium stretches=true, 
trapezium left angle=70, 
trapezium right angle=110, 
minimum width=3cm, 
minimum height=1cm, text centered, 
draw=black, fill=blue!30]

\tikzstyle{process1} = [rectangle, 
minimum width=3cm, 
minimum height=1cm, 
text centered, 
text width=5cm, 
draw=black, 
fill=white!30]

\tikzstyle{process} = [rectangle, 
minimum width=3cm, 
minimum height=1cm, 
text centered, 
text width=3cm, 
draw=black, 
fill=white!30]

\tikzstyle{decision} = [diamond, 
aspect=2
minimum width=3cm, 
minimum height=2cm, 
text centered, 
text width=4cm, 
draw=black, 
fill=white!30]
\tikzstyle{arrow} = [thick,->,>=stealth]

\usepackage{amsmath,amsfonts,amssymb,amsthm, bm}

\begin{document}

\title{Voltage-Dependent Electromechanical Wave Propagation Modeling for Dynamic Stability Analysis in Power Systems}

\author{Somayeh Yarahmadi,~\IEEEmembership{Student Member,~IEEE}, Daniel Adrian Maldonado,~\IEEEmembership{Member,~IEEE}, Lamine Mili,~\IEEEmembership{Life Fellow,~IEEE}, Junbo Zhao,~\IEEEmembership{Senior Member,~IEEE}, and Mihai Anitescu, ~\IEEEmembership{Member, ~IEEE}}



\maketitle

\begin{abstract}

Accurate dynamic modeling of power systems is essential to assess the stability of electrical power systems when faced with disturbances, which can trigger cascading failures leading to blackouts. A continuum model proves to be effective in capturing Electromechanical Wave (EMW) propagation characteristics, including its velocity, arrival time, and deviations. Analyzing these characteristics enables the assessment of the impacts of EMW on the performance of the protection system. Prior research has often modeled nonlinear EMW propagation through Partial Differential Equations (PDEs) within a homogeneous and uniform frame structure, assuming constant bus voltages across the entire power system. However, this assumption can produce inaccurate results. In this paper, we relax this assumption by introducing a second-order nonlinear hyperbolic EMW propagation equation model that accounts for voltage variations. Additionally, we present numerical solutions for the EMW propagation equation using the Lax-Wendroff integration method. To validate our approach, we conduct simulations on two test systems: a two-bus one-machine system and the New England 39-bus 10-machine system. The simulation results demonstrate the effectiveness of our proposed model and emphasize the importance of including the bus voltage equations in the analysis.
\end{abstract}


\section{Introduction}

\IEEEPARstart{U}nexpected events in a power system, such as sudden load changes, loss of generating units, and transmission system outages due to short circuits, result in an imbalance between the mechanical power input of generators and their electrical power output. This imbalance, known as accelerating power, can lead to loss of generator synchronization \cite{cresap}\cite{yan}. Specifically, changes in the generator rotor phase angles propagate through the entire power grid, a phenomenon known as EMW propagation, at velocities much slower than the speed of electromagnetic waves. However, these velocities increase when the total moment of inertia decreases, as observed in cases with a large penetration of renewable energy resources \cite{cui}. Therefore, it is vital that the protection systems react faster when this renewable energy penetration reaches a critical level and the probability of cascading failures becomes high \cite{Lei}. It has been shown that proper tuning of these relays requires accurate modeling of EMW propagation across the power grid, since the latter can induce mode instabilities and loss of synchronism \cite{ahad2015,parashar2003}.

In 1974, Semlyen \cite{semlyen} proposed an EMW propagation model by means of a continuous homogeneization approach. Semlyen's model consisted of a linear hyperbolic PDE and was used to analyze the impact of disturbances on the stability of the power system. Later, Thorp $et$ $al.$ \cite{thorp1998} developed a continuum model for the power system in 1-D and 2-D space and derived a nonlinear PDE to describe disturbance propagation. Following this work, Parashar $et$ $al.$  \cite{pashar2004} analyzed the EMW propagation using a pseudo-steady-state approach under the assumption of constant bus voltage magnitudes across the system while the bus voltage phase angles vary. In addition, they found that the parameters of the PDEs are anisotropic and nonhomogeneous. 

Bi $et$ $al.$ \cite{bi2017} improved the accuracy of the EMW propagation model using a structural frame model based on the distribution of the moment of inertia of synchronous generators through the power transmission network. They introduced an event-based protection system that responded to specific power system events by tailoring protective actions to the analysis of the EMW propagation. As a follow-up,  Huang $et$ $al.$ \cite{bi2020} proposed a nonhomogeneous continuum model that distributes the moment of inertia using the Gaussian function to account for the ohmic losses in the system.  However, this approach utilizes a simplified model where line resistances and voltage variations across the system are neglected. Using that approach, Dengyi $et$ $al.$ \cite{bi2019} investigated the impacts of the turbine-generator inertias, the line reactances, the bus voltages and the disturbance source frequencies on the propagation of the EMW and provided analytical solutions for each case to analyze the impact of the disturbance on the dynamics of the system. However, the assumptions considered, such as a sinusoidal disturbance with a specific frequency and magnitude and constant bus voltages, make the case studies unrealistic. 

Some applications of EMW propagation modeling considered in the literature include disturbance localization \cite{qin} \cite{es}, coherency identification \cite{slow}, and disturbance arrival time estimation \cite{rudez}. Few papers explored the impacts of the HVDC \cite{you2018}, high penetration of renewable energy resources \cite{cui}, and double-fed induction generators \cite{DFIG} on EMW propagation. Due to the converter interfaces of these devices, the total power system inertia is reduced and, thereby, the EMW propagation speeds up. Therefore, a better model of EMW propagation should be developed to properly tune the relays and mitigate the risk of cascading failures leading to blackouts.

To address this need, our paper develops an EMW propagation model that incorporates dynamic changes in voltage magnitudes and phase angles resulting from disturbances. In this model, we distribute the moments of inertia of the synchronous generators according to their potential energy and the admittances of the transmission lines, as described in \cite{Somayeh}. Although Qin $et$ $al.$ \cite{qin} presented an admittance-based inertia distribution that is limited to the reduced model of the simplified power system where only generator buses are considered and non-generator buses are removed during simplification, the approach used in this paper is applicable to non-simplified power system models. The resulting power system continuum model is more realistic since the non-homogeneity (function of space coordinates) and the non-isotropic characteristics of the parameters (function of wave direction) are considered.  The derived PDEs are solved using the Lax-Wendroff integration method, which makes it possible to investigate disturbances due to short circuits or changes in the load or generation  instead of considering only sinusoidal disturbances. Notably, since this paper assesses the dynamics of the voltages in the power system and the voltages are parameters of the EMW velocity, which are functions of their location, we name the voltage-dependent and voltage-independent EMW analysis as follows: EMW homogeneous and EMW nonhomogeneous modeling, respectively. Numerical simulations performed on various power test systems show that variations in bus voltage magnitudes have an important impact on EMW velocities and arrival times. It is important to note that AVRs and PSSs were not included in the scope of this study. In future research, we will investigate the impacts of AVRs and PSSs on EMW propagation. Additionally, this paper exclusively focuses on the analysis of EMW propagation and does not involve any protection-related studies.

\color{black}

The paper is organized as follows. Section II describes the mechanical analogy of the power system in EMW analysis. Section III studies a homogeneous EMW propagation modeling with a constant bus voltage assumption. Section IV derives the nonhomogeneous EMW propagation that accounts for bus voltage changes in a power system subjected to a disturbance. Section V presents a numerical method to solve the proposed non-homogeneous equations. Section VI considers several case scenarios to examine the proposed EMW equation and discusses the simulation results. Finally, Section VII concludes the paper and presents future research work.

\textit{\textbf{Remark}: In this paper, all powers, currents, and voltages are functions of $\xi$ and $t$ except in certain cases that are indicated otherwise.}

\section{Mechanical analogy of the power system components}

In the realm of EMW analysis within power systems, the mechanical analogy of a transmission line serves as a valuable conceptual framework, drawing parallels between the behavior of electrical transmission lines and mechanical systems. By likening electrical variables to mechanical quantities such as force, velocity, and impedance, we gain a deeper understanding of how waves propagate through power networks.

Consider a rod with length denoted L. In structural mechanics, torsional motion manifests itself as a propagation of a twisting disturbance along the rod's longitudinal axis. Similarly, in power systems, various parameters find analogous counterparts that contribute to our understanding of EMWs and their impact on grid stability and resilience. The rotating machinery, including generators and turbines in a power system, finds an analogy in mass density ($\rho$) within the concept of mechanical waves, which affects their response to external forces. When encountering higher inertia, analogous to higher mass density, these components demonstrate greater resistance to rapid alterations in speed or phase angles during disturbances, significantly contributing to grid stability. Furthermore, the moment of inertia of a synchronous generator rotor corresponds to the polar moment of inertia (I) in the mechanical system, influencing the resistance of the rotor to changes in speed or phase angle, thus preventing disruptive behavior during transient events. Another parameter in the power system, the electrical stiffness of transmission lines, and its analogous counterpart, the field winding stiffness of synchronous generators represented by the shear modulus (G), determines the system's capacity to resist significant phase deviations during disturbances, enhancing phase stability within the grid and contributing to reliable power transmission. Moreover, angular displacement represents the phase angle difference between various grid components, reflecting electrical phase shifts during EMW events. Accurate monitoring and understanding of this phase angle difference are crucial for assessing power system stability and behavior under disturbances, enabling informed decision making by engineers and grid operators. 

By extending these analogies into the realm of power systems, we equip ourselves with a robust framework for understanding and predicting the intricate behavior of EMWs, valuable tools in the design, analysis, and optimization of power grids, ensuring reliability, stability, and resilience of the electrical system\cite{bi2017}, \cite{somayeh2023}, and \cite{pashar2004}. Table \ref{table:sth} illustrates the parameters of the power system and their mechanical analogues in the studies of torsional waves.

\begin{table}[!hbt]
\centering
\caption{Analogy between mechanical and electrical parameters}
\renewcommand{\arraystretch}{1.5} 
\begin{tabular}{|c | c|} 
 \hline  
\textbf{Mechanical parameters} & \textbf{Electrical Parameters} \\
\hline
Mass density ($\rho$) & Inertia of rotating machinery\\ & (including generators and turbines) \\
\hline
Polar moment of inertia (\(I\)) & Moment of inertia of generator rotor ($J$)\\
\hline
Shear modulus (\(G\)) & Electrical stiffness of transmission lines \\ & (electrical impedance)\\
\hline
Angular displacement ($\Delta \theta$) & Phase angle difference $\Delta \delta$\\
\hline 
\end{tabular}
\label{table:sth}
\end{table}

\section{Homogeneous EMW propagation Model}
This section describes the Admittance-Based Inertia Distribution (ABID) method employed to implement the EMW propagation in both homogeneous and nonhomogeneous modeling. Then homogeneous EMW propagation modeling is derived as a linear hyperbolic EMW equation. Finally, the EMW propagation features are discussed.

\subsection{The Moment of Inertia Distribution} \label{sec:2.1a}
The electromagnetic energy derives from the electromagnetic waves propagating through the transmission lines, while the EMW  energy derives from an exchange between the kinetic energy stored in the rotor's rotation and the potential energy stored in the line admittances and shunt susceptances. The conventional power system model involving differential-algebraic equations does not account for the EMW energy and associated EMW wave propagation through the transmission network. To model the EMW propagation, a continuum modeling is suggested that considers the parameters of the power system components such as generator inertias and resistances, transmission line reactances, resistances and shunt capacitances, and loads distributed along the transmission lines \cite{es}.

\begin{figure}[htb!]
    \centering
    \includegraphics[scale=0.3]{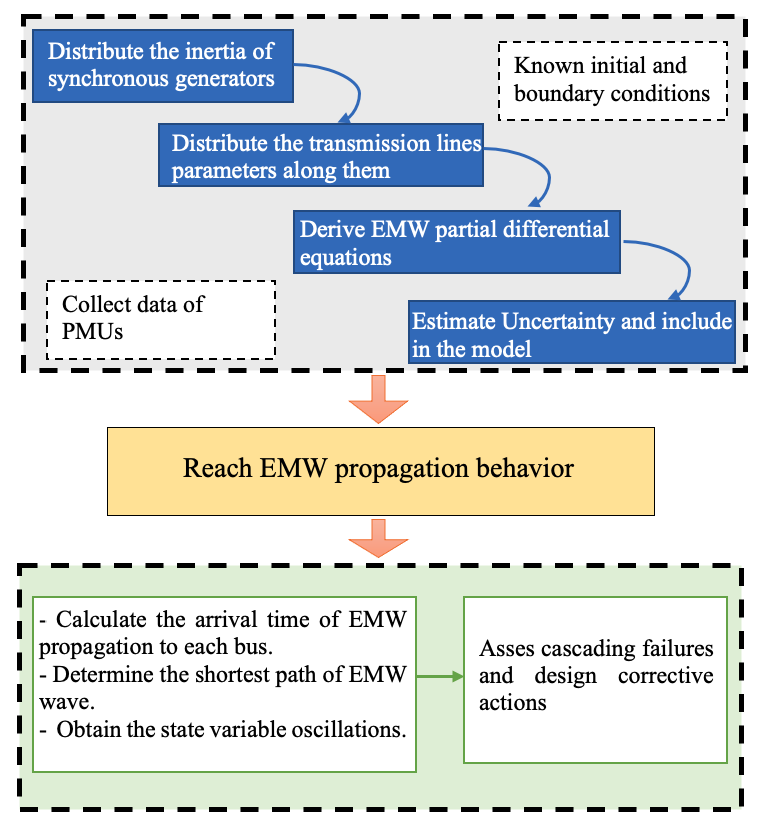}
    \caption{Outline of the EMW propagation analysis framework.}
    \label{fig:p}
\end{figure}

Figure \ref{fig:p} displays the EMW analysis framework, which consists of three main steps: 1) developing the rotor angle deviation equations; 2) finding a solution to the EMW equations; and 3) analyzing the EMW behavior. A continuum model is employed to derive the EMW wave equations. To this end, the inertia of the synchronous machines and the line parameters are distributed throughout the transmission network.  Accordingly, the EMW partial differential equations can be conducted using the power flow and swing equations of the infinitesimal components of the transmission lines.  As for the EMW velocity, arrival time, and other EMW characteristics, they are calculated from the EMW equations. Once the latter are numerically solved, we can initiate corrective actions aimed at protecting the power system from instability and cascading failures that may lead to blackouts.

In this paper, we employ the approach proposed in \cite{Somayeh}, called the ABID approach. In this approach, when a disturbance occurs in a power system, the transient potential energy of a transmission line changes proportionally with the line admittance \cite{cai2007}. Specifically, ABID consists of two main steps:
\begin{itemize}
    \item Step1: the inertia of a generator is distributed between its incident lines proportionally to their admittances;
    \item Step 2: if a line with inertia $j_i$  and admittance $Y_i$ is connected to a nongenerator bus, the associated moment of inertia changes to $ \frac{n Y_i}{n  Y_i + \sum_{j=1}^{n} Y_j} j_i$ and the difference will distribute to the $n$ connected  lines to the non-generator bus.
\end{itemize}   
To distribute the inertia of a generator across the transmission network, these two steps are repeated until all generator buses are attained.

\subsection{Homogeneous EMW Propagation Equation}
Let us assume that the generators are uniformly distributed along the transmission lines. In that case, the active power change along a very small part of a line can be found via   
\begin{align}
    & \Delta P = \frac{\partial P_t}{\partial \xi} \Delta \xi, \label{eq:deltap1_hom}
\end{align}
and using the swing equation, this power is equal to 
\begin{align}
      & \Delta P  = - j_h \omega_0 \frac{\partial ^2 \delta }{\partial ^2t} \Delta \xi, \label{eq:deltap2_hom}
\end{align}
where $P_t$, $j_h$, and $\omega_0$ are the power distribution in a line in MW, the moment of inertia per unit length in kg-m$^2$, and the  synchronous speed in rad/s, respectively. 

\color{black}
The power transmitted over a $\Delta \xi$ length of a line is given by
\begin{align}
    & P_t = \frac{V^2}{x \Delta \xi } sin (-\Delta \theta) = V^2 b \frac{sin (-\Delta \theta)}{\Delta \xi },  
\end{align}
where, $b = 1/x$, and $V$ and $x$ are the voltage and the reactance per unit length of the line, respectively. For $\Delta \xi \approx 0$, we approximate $sin (\Delta \theta) \approx \Delta \theta$. Therefore, we get
\begin{align}
    & P_t = -V^2 b \frac{ \Delta \theta}{\Delta \xi } = V^2 b \frac{\partial \theta}{\partial \xi }\label{eq:pt_hom}.
\end{align}
Substituting (\ref{eq:pt_hom}) in (\ref{eq:deltap1_hom}) and using (\ref{eq:deltap2_hom}), yields 
\begin{align}
  & \frac{\partial ^2 \theta}{\partial t^2} =  (\frac{V^2 b}{j_h \omega_0}) \frac{\partial ^2 \theta}{\partial \xi^2} \label{eq:emw_hom}.
\end{align}
This is a second-order linear hyperbolic wave equation that is presented in \cite{bi2017}\cite{semlyen} as the EMW propagation equation. It is based on the following assumptions:
\begin{itemize}
    \item constant voltage in the entire power system;
    \item neglecting the damping torque, inductance, and resistance of the synchronous generators;
    \item ignoring resistance and susceptance of lines;
    \item ignoring non-homogeneous and anisotropic characteristics of parameters of the power system.
\end{itemize}
Under these assumptions, we can calculate the EMW velocity using (\ref{eq:emw_hom}),  which is equal to $\sqrt{\frac{V^2 b}{j_h \omega_0}}$. The velocity rises as $j_h$ decreases. This is why high penetration of renewable energy, normally with low inertian response, increases the EMW velocity.

\section{Proposed Non-homogeneous EMW Propagation modeling}

When a power system experiences stressed conditions, the coupling between the voltage and rotor angle stability is strong, which are both affected by active and reactive power flows \cite{kundur}. Therefore, the dynamical behavior of the bus voltages impacts the EMW propagation. This section provides nonhomogeneous disturbance propagation modeling while considering voltage changes in the power system. Also, it derives a partial differential equation that accounts for voltage variations along the transmission lines.

\subsection{Power Flow Distribution Across the Transmission Lines}
An active power flow distribution is needed to achieve a continuous EMW propagation model. Therefore, we first develop the power flow equations for a line with lumped parameters. Then, we use those power flow equations to derive a continuous power flow distribution for a line with infinitesimal components.  
 
 Consider a lumped $\pi$ equivalent model of a transmission line, where $R$, $X$, and $B_c$ are, respectively, the resistance, reactance and shunt susceptance of the line. The transferred power through that line can be expressed as
\begin{align}
     P+jQ &= \overline{V}_1 \overline{I}_1^* =\overline{V}_1 (\frac{\overline{V}_1 - \overline{V}_2}{Z})^* +  \overline{V}_1 (j \overline{V}_1 \frac{B_c}{2})^* \nonumber \\
    & = \overline{V}_1 (\overline{V}_1 -  \overline{V}_2)^*(G+jB) - j V_1^2 \frac{B_c}{2},
\end{align}
where $\frac{1}{Z^*} = G+jB$ and $\overline{V}_1 = V_1 e^{\theta _1}$ and $\overline{V}_2 = V_2 e^{\theta _2}$ are the phasor voltages of the two ends of the line. Note that $B_c$ only contributes to the reactive power. Therefore, since the goal is to derive the active power flow equation, $B_c$ is ignored when calculating the transmitted power.

Let $\Delta S$ be the apparent power deviation at $\xi$. Performing a forward and backward of EMW propagation results in $\Delta S$ being equal to the sum of the apparent power flow from $\xi$ to $\xi_+$ and the apparent power flow from $\xi$ to $\xi_-$.

\medskip

Formally, we have
\begin{align}
      \Delta S = S_+ +  S_- = & \overline{V} (\overline{V}-  \overline{V}_{+})^*(g+jb)\frac{1}{\Delta \xi} \nonumber \\ 
      & + \overline{V} (\overline{V} -  \overline{V}_{-})^*(g+jb)\frac{1}{\Delta \xi}, \label{eq:delp3}
\end{align}
where $\Delta \xi$, $\overline{V}$, $\overline{V}_{-}$, $\overline{V}_{+}$ are length of the component and voltage at $\xi$, $\xi_-$, and $\xi_+$, respectively and $g = Gl$ and $b = Bl$ that $l$ is length of the line.

Using a second-order Taylor series expansion at $\xi$ for  ${\xi_+}$ and ${\xi_-}$ yields
\begin{align}
    \overline{V}_{+} = \overline{V} + \frac{\partial \overline{V}}{\partial \xi} \Delta \xi + \frac{1}{2} \frac{\partial ^2 \overline{V}}{\partial \xi^2} \Delta \xi^2 + ..., \label{eq:taylorvplus} \\
    \overline{V}_{-} = \overline{V} - \frac{\partial \overline{V}}{\partial \xi} \Delta \xi + \frac{1}{2} \frac{\partial ^2 \overline{V}}{\partial \xi^2} \Delta \xi^2 + ..., \label{eq:taylorvminus} 
\end{align}
Substituting (\ref{eq:taylorvplus}) and (\ref{eq:taylorvminus}) in (\ref{eq:delp3}) yields
 \begin{align}
       \Delta S = & - \overline{V} (\frac{\partial ^2 \overline{V}}{\partial \xi^2})^*(g+jb)\Delta \xi. \label{eq:deltas}
\end{align}
Let us now derive the expression of $\frac{\partial ^2 \overline{V}}{\partial \xi^2}$. We have 
\begin{align}
    & \frac{\partial \overline{V}}{\partial \xi} = \frac{\partial V}{\partial \xi}e^{j\theta} + j \frac{\partial \theta }{\partial \xi}V e^{j\theta}, \\
    & \frac{\partial ^2 \overline{V}}{\partial \xi^2} =  \frac{\partial}{\partial \xi}[\frac{\partial V}{\partial \xi}e^{j\theta} + j \frac{\partial \theta }{\partial \xi}V e^{j\theta}] = \nonumber\\
     &= [\frac{\partial^2 V}{\partial \xi^2} - (\frac{\partial \theta }{\partial \xi})^2 V] e^{j\theta} + j[ \frac{\partial^2 \theta}{\partial \xi^2} V  + 2 \frac{\partial V}{\partial \xi}\frac{\partial \theta }{\partial \xi}]e^{j\theta}. \label{eq:dif2V}
\end{align}
Substituting (\ref{eq:dif2V}) in (\ref{eq:deltas}) yields
\begin{align}
 \Delta S &  = \Delta P + j \Delta Q =  -([V \frac{\partial^2 V}{\partial \xi^2} - V^2 (\frac{\partial \theta }{\partial \xi})^2]  - j[ V^2 \frac{\partial^2 \theta}{\partial \xi^2}  \nonumber \\
&+ 2 V \frac{\partial V}{\partial \xi}\frac{\partial \theta }{\partial \xi}])(g+jb)\Delta \xi .
\end{align}
Therefore, $\Delta P$ and $\Delta Q$ are given by
\begin{align}
 \Delta & P =&\\
 &-g\Delta \xi[V \frac{\partial^2 V}{\partial \xi^2} - V^2 (\frac{\partial \theta }{\partial \xi})^2] - b \Delta \xi [ V^2 \frac{\partial^2 \theta}{\partial \xi^2}  + 2 V  \frac{\partial V}{\partial \xi}\frac{\partial \theta }{\partial \xi}], \label{eq:eqdeltap1}\\
 \Delta & Q =&\\
&g \Delta \xi [ V^2 \frac{\partial^2 \theta}{\partial \xi^2}  + 2 V \frac{\partial V}{\partial \xi}\frac{\partial \theta }{\partial \xi}] - b \Delta \xi[V \frac{\partial^2 V}{\partial \xi^2} - V^2 (\frac{\partial \theta }{\partial \xi})^2] \label{eq:eqdeltaq1}.
\end{align}
The equations (\ref{eq:eqdeltap1}) and (\ref{eq:eqdeltaq1}) respectively prove the expressions of $\Delta P$ and $\Delta Q$ while the voltage amplitude and the voltage angle vary along a transmission line. These equations show that both active and reactive power flows depend on voltage and angle deviations.

\subsection{Nonhomogeneous EMW Equation}
In this section, the line Ohmic losses, $g$, are assumed to be negligible. Thus, we have
\begin{align}
& \Delta P =  -[ V^2b \frac{\partial^2 \theta}{\partial \xi^2}  + 2 V b \frac{\partial V}{\partial \xi}\frac{\partial \theta }{\partial \xi}] \Delta \xi \label{deltap},\\
& \Delta Q =  -[Vb \frac{\partial^2 V}{\partial \xi^2} - V^2b (\frac{\partial \theta }{\partial \xi})^2] \Delta \xi. \label{deltaq}
\end{align}
In order to have a continuum model, the moment of inertia of the synchronous generators should be distributed throughout the transmission network. As mentioned, we employ the inertia distribution method proposed in \cite{Somayeh}. Thus, the swing equation for a segment of a line of length of $\Delta \xi$  is given by
\begin{align}
   j_h \Delta \xi \frac{\partial \theta^2}{\partial t^2} = -\Delta P \label{eq:j},
\end{align}
where $j_h$ is the moment of inertia per unit length of the line. Substituting (\ref{deltap}) into (\ref{eq:j}), yields
\begin{align}
    j_h  \omega_0 \frac{\partial ^2 \theta}{\partial t^2} =   V^2b \frac{\partial^2 \theta}{\partial \xi^2} + 2 V b \frac{\partial V}{\partial \xi}\frac{\partial \theta }{\partial \xi}. \label{eq:0202}
\end{align}
(\ref{eq:0202}) shows the nonhomogeneous EMW equation while the magnitude and the angle of the bus voltages have dynamic behaviors. 

If we assume that the voltage does not change with respect to $\xi$, (\ref{eq:0202}) reduces to
\begin{align}
    j_h \omega_0 \frac{\partial ^2 \theta}{\partial t^2} =   V^2 b\frac{\partial ^2 \theta}{\partial \xi^2}. \label{eq:0020}
\end{align}
This is the homogeneous EMW equation presented in (\ref{eq:emw_hom}).

\subsection{Voltage equation}
The nonhomogeneous EMW propagation model can be rewritten as
\begin{align}
    & \frac{\partial \Delta \theta}{\partial t} = \Delta \omega \label{eq:EMW1},\\
    & \frac{\partial  \Delta \omega}{\partial t} = - \frac{1}{j \omega_0 }\frac{\partial P}{\partial \xi} \label{eq:EMW2},\\
    & P = - V^2 b \frac{\partial \Delta \theta}{\partial \xi}. \label{eq:EMW3}
\end{align}
Note that for the homogeneous EMW propagation model, the voltage is assumed to be constant in (\ref{eq:EMW3}).

The set of EMW propagation equations ((\ref{eq:EMW1}), (\ref{eq:EMW2}) and (\ref{eq:EMW3})) shall be supplemented by an additional equation describing how the voltage varies in space. To this end, for a single one-dimensional line, we discretize the line in $n + 1$ points $\{\xi_0, \xi_1, . . ., \xi_k, . . . , \xi_n\}$. This gives us a set of discretized voltages $\{v_0, v_1, . . .,v_k, . . . , v_n\}$. $v_0$ and $v_n$ are the voltages at boundaries and $v_k, _{k\neq \{0,n\}}$ are the voltages at the interior line's points. The discretized voltages in the interior ($\{v_1, . . . , v_{n-1} \}$) can be obtained by writing the Kirchoff's equations as (\ref{eq:mtrx}), where $g_0(t)$ and $g_n(t)$ are the boundary voltage conditions at buses that are defined by boundary equations at the buses.  For instance, if that boundary is connected to a synchronous generator, the exciter determines the voltage of that bus.$y_{i,i}$ represents the shunt admittance connected to point $i$, while $y_{i,j}$ denotes the admittance of the component of the line between point $i$ and its neighboring point $j$.

\begin{align} \label{eq:mtrx}
\begin{bmatrix}
    1\ ...  &0& 0& 0 &...\ 0\\
 &  & \vdots & &\\
    0\ ... &-y_{k,{k-1}}& y_{k,{k-1}}+y_{k,{k+1}} & -y_{{k},{k+1}} &...\ 0\\
     &  & \vdots & &\\
     0\ ... & 0& 0&0&...\ 1
\end{bmatrix} \times  \nonumber\\ \begin{bmatrix}
    v_0(t)\\
    \vdots\\
    v_{k-1}(t)\\
    v_k(t)\\
    v_{k+1}(t)\\
    \vdots\\
    V_n
\end{bmatrix}= \begin{bmatrix}
    g_0(t)\\
    \vdots\\
    0\\
    0\\
    0\\
    \vdots\\
    g_n(t)
\end{bmatrix}.
\end{align}

At the interior points, we have
\begin{equation}
    -y_{k,{k-1}}v_{k-1} + (y_{k,{k-1}}+y_{k,{k+1}}) v_{k} - y_{{k},{k+1}} v_{k+1} = 0.
\end{equation}
The line admittance is assumed to be homogeneous and is discretized uniformly. Therefore, we have 
\begin{equation}
    b_k(v_{k-1} - 2 v_{k} +  v_{k+1}) = 0.
\end{equation}
Kirchoff's equations, which can be written with the incidence matrix and the primitive admittance matrix ($Y V = B^T Y_p B = I$), stand for the discretized version of the Laplace operator. This leads to
\begin{equation} \label{eq:eqv2}
    \frac{\partial^2 V}{\partial^2 \xi} = 0.
\end{equation}

\section{Numerical Integration of EMW Propagation Equation}
The EMW propagation equation is a hyperbolic PDE with finite wave propagation. Therefore, we resort to the Richtmyer two-step Lax–Wendroff method to solve it.

\subsection{The Lax–Wendroff Method}
The Lax-Wendroff method is a numerical approach proposed by Lax and Wendroff \cite{lax} to solve nonlinear hyperbolic partial differential equations based on finite differences.  Consider a first-order PDE with two variables expressed as

\begin{equation}
    \frac{\partial u(\xi,t)}{\partial t} + \frac{\partial f(u(\xi,t))}{\partial \xi} = 0 \label{eq:lax},
\end{equation}
where f is a function of $\xi$ and $t$, which are independent variables. In the linear case given by $f(u(\xi,t)) = A u(\xi,t)$, a numerical solution of (\ref{eq:lax}) is given by

\begin{equation}
    u_i^{k+1} = u_i^{k} - \frac{\Delta t}{2 \Delta \xi} A [u_{i+1}^{k} - u_{i-1}^{k}] + \frac{\Delta t^2}{2 \Delta \xi^2} A^2 [u_{i+1}^{k} - 2u_i^{k} + u_{i-1}^{k}], \label{eq:3111}
\end{equation}
where $k$ and $i$ refer to $\xi$ and $t$, respectively.
If the f is nonlinear, the numerical solution will be

\begin{align}
    & u_i^{k+1} = u_i^{k} - \frac{\Delta t}{2 \Delta \xi}[f(u_{i+1}^k) - f(u_{i-1}^k)] + \nonumber\\
    & \frac{\Delta t^2}{2 \Delta \xi}[C_{{i+1}/2}(f(u_{i+1}^k) - f(u_i^k)) - C_{{i-1}/2}(f(u_i^k) - f(u_{i-1}^k))], \label{eq:3222}
\end{align}
where $C_{i\pm 1/2}$ is the Jacobian matrix evaluated at $\frac{1}{2}(u_i^k + u_{i\pm 1}^k)$. Some methods such as the MacCormack method and Richtmyer two-step Lax–Wendroff technique avoid calculating the Jacobian matrix. For instance, for the Richtmyer method, the steps are as follows:

\begin{itemize}
    \item Step one: Solving the equation for half step in time.
    
    \begin{align}
    & u_{{i+1}/2}^{{k+1}/2} = \frac{1}{2} (u_{i+1}^k + u_i^{k}) - \frac{\Delta t}{2 \Delta x} (f(u_{i+1}^k) - f(u_{i}^k)),\\
    & u_{{i-1}/2}^{{k+1}/2} = \frac{1}{2} (u_{i}^k + u_{i-1}^{k}) - \frac{\Delta t}{2 \Delta x} (f(u_{i}^k) - f(u_{i-1}^k)).
\end{align}
    \item Step two: Finding $u_i^{k+1}$

    \begin{equation}
    u_i^{k+1} = u_i^{k} - \frac{\Delta t}{\Delta x} [f(u_{{i+1}/2}^{{k+1}/2}) - f(u_{{i-1}/2}^{{k+1}/2})].
\end{equation}
\end{itemize}

\subsection{Numerical Solution of the EMW Propagation Equations}
As it was discussed, the EMW propagation equation and the voltage equation along a line, respectively, are given by

\begin{align}
      & \frac{\partial ^2 \Delta \theta}{\partial t^2} =  \nu^2 V^2 \frac{\partial ^2 \Delta \theta}{\partial \xi^2} + 2 \nu^2 V \frac{\partial V}{\partial \xi} \frac{\partial \Delta \theta}{\partial \xi},\\
      & \frac{\partial^2 V}{\partial \xi^2} = 0 \label{eq:vv},
\end{align}
where $\nu^2 = \frac{b}{j_h}$. To convert the second-order PDE to a first-order PDE, which is solvable by the Lax-Wendroff equation, let us define $ \chi = \Delta \omega= \frac{\partial \Delta \theta}{\partial t}$, $\lambda = -\nu \frac{\partial \Delta \theta}{\partial \xi}$, and $\gamma = -\nu V^2 \frac{\partial \Delta \theta}{\partial \xi}$. Therefore, we have

\begin{align}
      &  \frac{\partial \lambda}{\partial t} = - \nu \frac{\partial \chi}{\partial \xi}, \label{eq:377}\\
      & \frac{\partial \chi}{\partial t} =  - \nu \frac{\partial \gamma}{\partial \xi},  \label{eq:388}\\
      & \gamma = V^2 \lambda.
\end{align}
\textit{Initial and boundary conditions:} In order to find the numerical solution of these equations, firstly we need to define the boundary and the initial conditions at $\xi=0$ and $t=0$. These conditions are given by
\begin{align*}
    & \Delta \theta(\xi,t)|_{\xi = 0} = \Delta \theta(0,t), \\
    & \chi|_{\xi = 0} = \Delta \omega|_{\xi = 0}  = \frac{\partial \Delta \theta(\xi,t)}{\partial t}|_{\xi = 0} ,\\
    & \gamma|_{\xi = 0} =- V^2 \nu \frac{\partial \Delta \theta(\xi,t)}{\partial \xi}|_{\xi = 0} = b \nu \Delta P|_{\xi = 0}, \\
    & V(\xi,t)|_{\xi =0} = V(0,t).
\end{align*}
To update the boundary value for $\Delta \theta$, we interpolate the equations in the following way:
\begin{itemize}
    \item first, we update the values along the branch, using the boundary values of the previous time step;
    \item second, after the values are updated, we update the boundary values of $\Delta \theta$ by copying the value of the contiguous variable.
\end{itemize}
\textit{Finite differences:} To solve the EMW equations using the Lax-Wendroff method, we calculate the finite differences for (\ref{eq:377}) and (\ref{eq:388}). Note that here we only provide first-order differences; (\ref{eq:3111}) and (\ref{eq:3222}) show how to include second-order differences. The finite differences are given by
\begin{align}
    \frac{\lambda_i^{k+1} - \lambda_i^{k}}{\Delta t} = - \nu \frac{\chi_{i+1}^{k} - \chi_{i-1}^{k}}{2\Delta \xi}, \label{eq:411}\\
    \frac{\chi_i^{k+1} - \chi_i^{k}}{\Delta t} = - \nu \frac{\gamma_{i+1}^{k} - \gamma_{i-1}^{k}}{2\Delta \xi}.\label{eq:422}
\end{align}
If we discretize (\ref{eq:vv}), we get
\begin{align}
    & v_{i}^{k+1} = 2 v_{i}^{k} - v_{i}^{k-1}. \label{eq:433}
\end{align}
Therefore, $\gamma_{i}^{k+1}$ can be calculated as
\begin{align}
    \gamma_{i}^{k+1} = {v_{i}^{k+1}}^2 \lambda_{i}^{k+1}.\label{eq:444}
\end{align}

Each line is discretized according to the spatial step size selected $\Delta \xi$, and its length. Each discretization point in the branch will have two degrees of freedom, $\chi$ and $\lambda$. 

For the inner points of the branch from 1 to $n - 2$, we write the finite centered differences, and for 0 and $n - 1$ the situation is not that clear because these points are at the boundary and we cannot use the centered differences. To solve this problem, \cite{Chaudhry} proposes two solutions, which are
\begin{itemize}
    \item use the characteristic equations at the boundary;
    \item extrapolate using fictitious grid points. Let $f_{n-1}$ be the flux at the boundary; we define $f_{n} = 2f_{n-1} - 2f_{n-2}$. Then, take a forward finite difference approximation.
\end{itemize}
As boundary conditions are given, we choose the first solutions for the start and end points of the lines.

Let us apply the numerical technique to calculate the EMW propagation in a power system from Generator $d$ ($G_d$) to Generator $q$ ($G_q$). After a disturbance occurs, power flow calculations are performed as the first step. The next steps are distributing the inertia of the synchronous generators and finding the short path from every generator to the other generators. Note that the shortest EMW propagation path can be found using the EMW velocity and applying Dijkstra's algorithm \cite{bi2017}. Then, the shortest EMW path between $G_d$ and $G_q$ is discretized starting from $G_d$ ($\xi=0$) is discretized and $\chi_i^{k+1}$, $\lambda_i^{k+1}$, $v_i^{k+1}$, $\gamma_i^{k+1}$, and $\Delta \theta_i^{k+1}$ are calculated under the assumption that their values are known at k=0.  This process is repeated until a bus is reached. Then if the bus is connected to a generator, say $G_q$, the calculation is stopped; otherwise, the initial and boundary conditions are updated based on the power flow at the bus, and the variables at $k+1$ are calculated for the new line.

\medskip

\section{Simulation Results}
In this section, we assess the performance of our proposed approach through simulations conducted on two distinct test systems: the 2-bus 1-machine system \cite{Bukhsh} and the New England 39-bus 10-machine system. Our primary objective is to investigate the propagation behaviors of EMW within these systems under various scenarios and system conditions. Additionally, we compare the results of the second test system with the results of the Power Factory (PF) simulation. The following subsections provide detailed information on the simulation methodology and the specific scenarios considered for each test system.

\subsection{The 2-Bus 1-Machine System}

The 2-bus 1-machine system, which serves as a simple power system in our study, comprises two interconnected buses connected by a transmission line with defined parameters: a line resistance ($r$) of 0.04 ohms and a reactance ($x$) of 0.2 ohms. At Bus 2, we have set fixed values for both active and reactive powers, with the active power fixed at 352.5 MW and the reactive power at -358 MVAr, respectively.

\begin{figure}
  \begin{subfigure}{\linewidth}
    \centering
    \includegraphics[width=.7\linewidth, height=120pt]{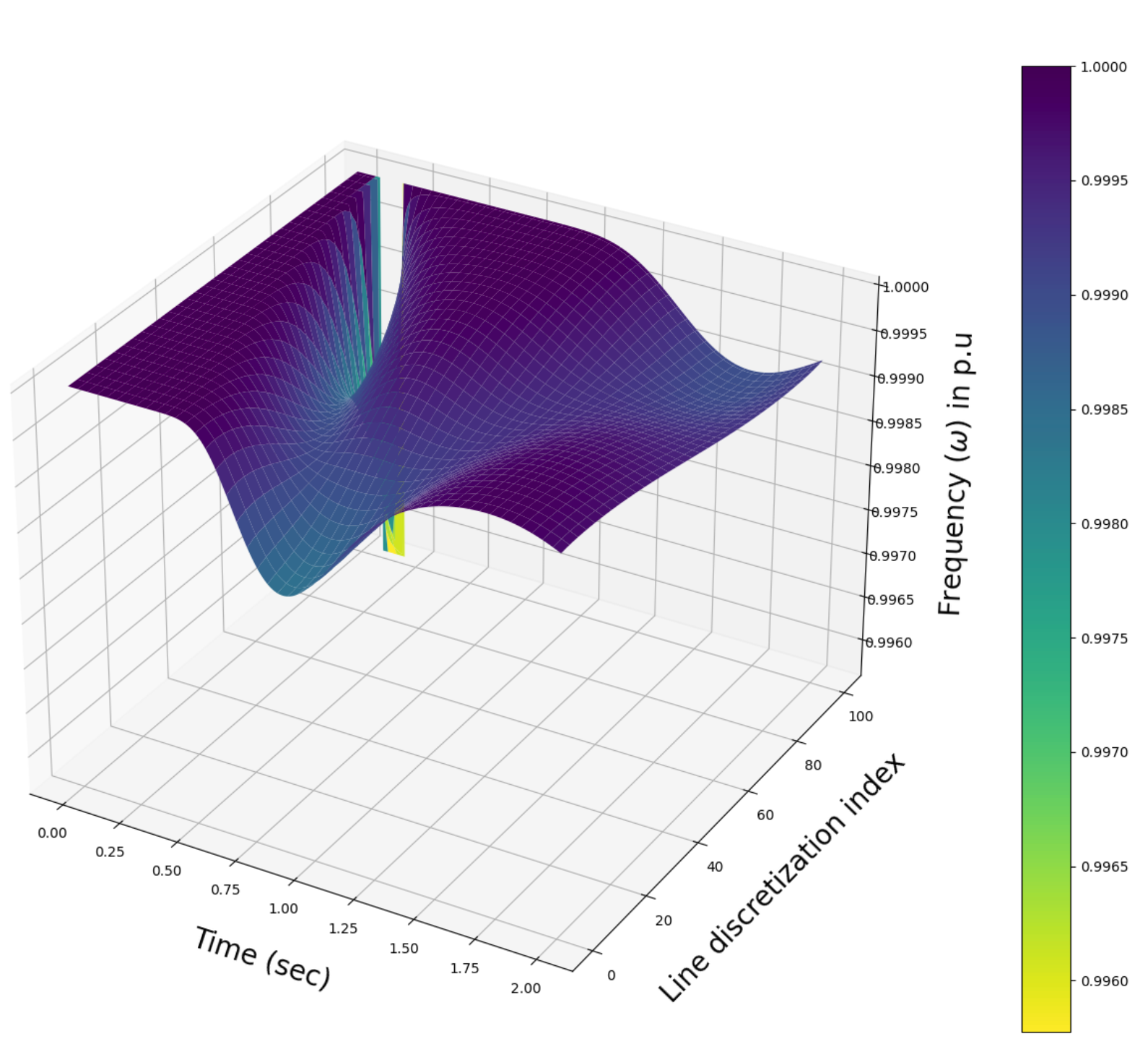}
    \caption{Rotor angular velocity (H = 1.5 MJ/MVA.)}
  \end{subfigure}

  \vspace{\floatsep}

  \begin{subfigure}{\linewidth}
    \centering
    \includegraphics[width=.7\linewidth, height=120pt]{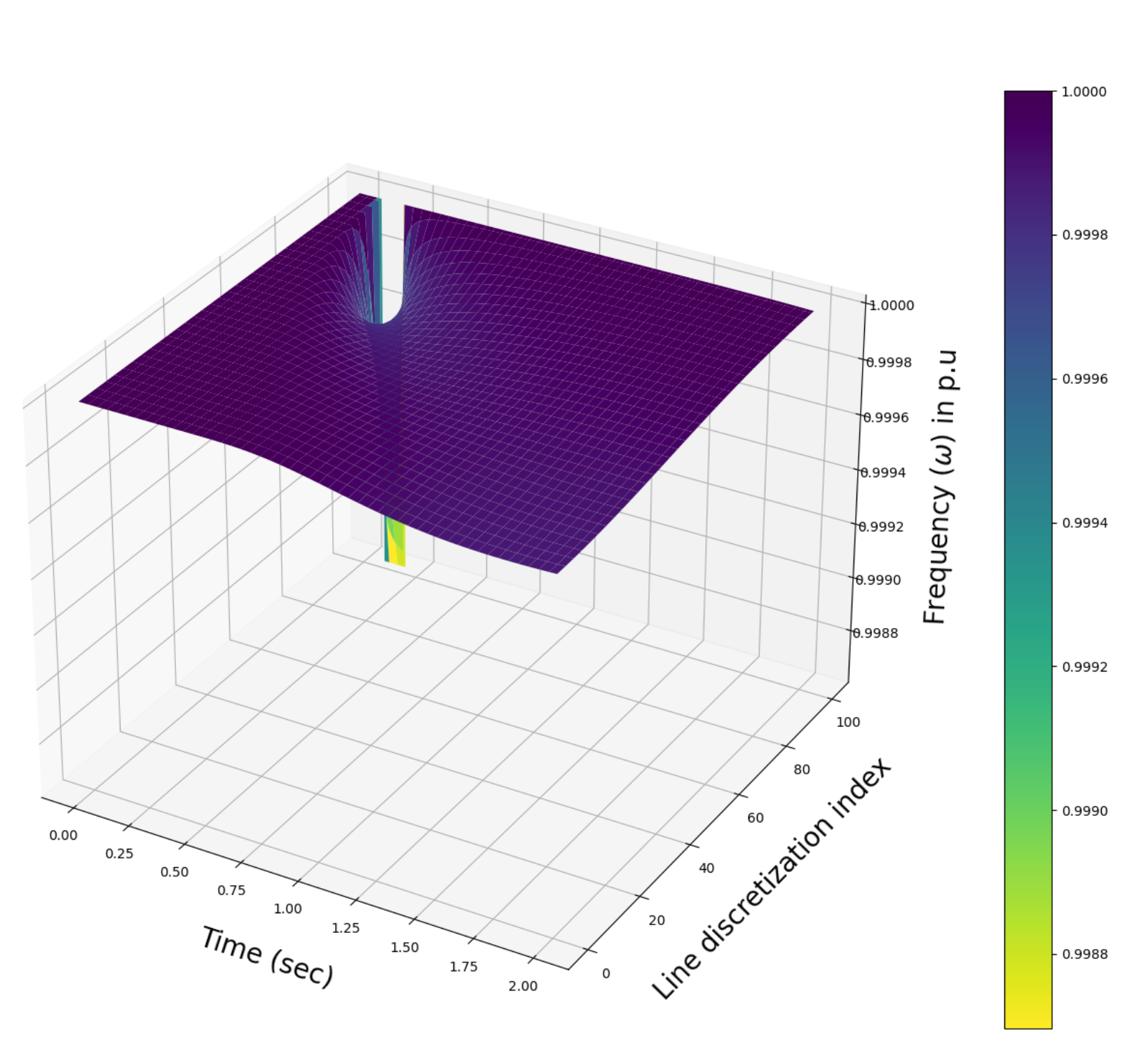}
    \caption{Rotor angular velocity (H = 15 MJ/MVA.)}
  \end{subfigure}

  \caption{Rotor angular velocity of the 2-bus power system induced by an increase in active load power by 10\%.}
  \label{fig:2bussystemresults}
\end{figure}

\begin{figure}
  \centering

  \begin{subfigure}{0.7\linewidth}
    \includegraphics[width=\linewidth, height=120pt]{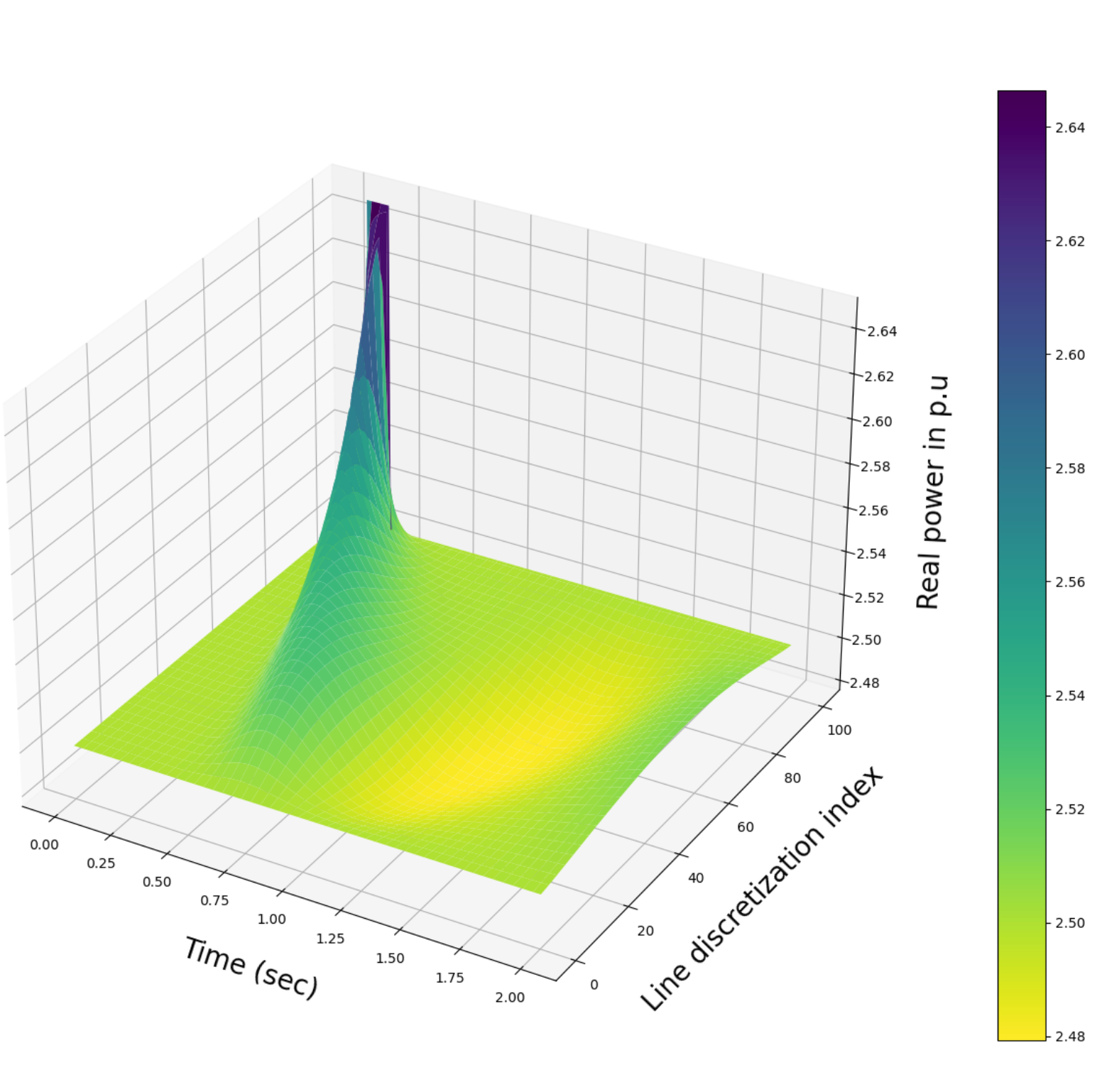}
    \caption{Electrical power response (H = 1.5 MJ/MVA.)}
  \end{subfigure}

  \vspace{\floatsep}

  \begin{subfigure}{0.7\linewidth}
    \includegraphics[width=\linewidth, height=120pt]{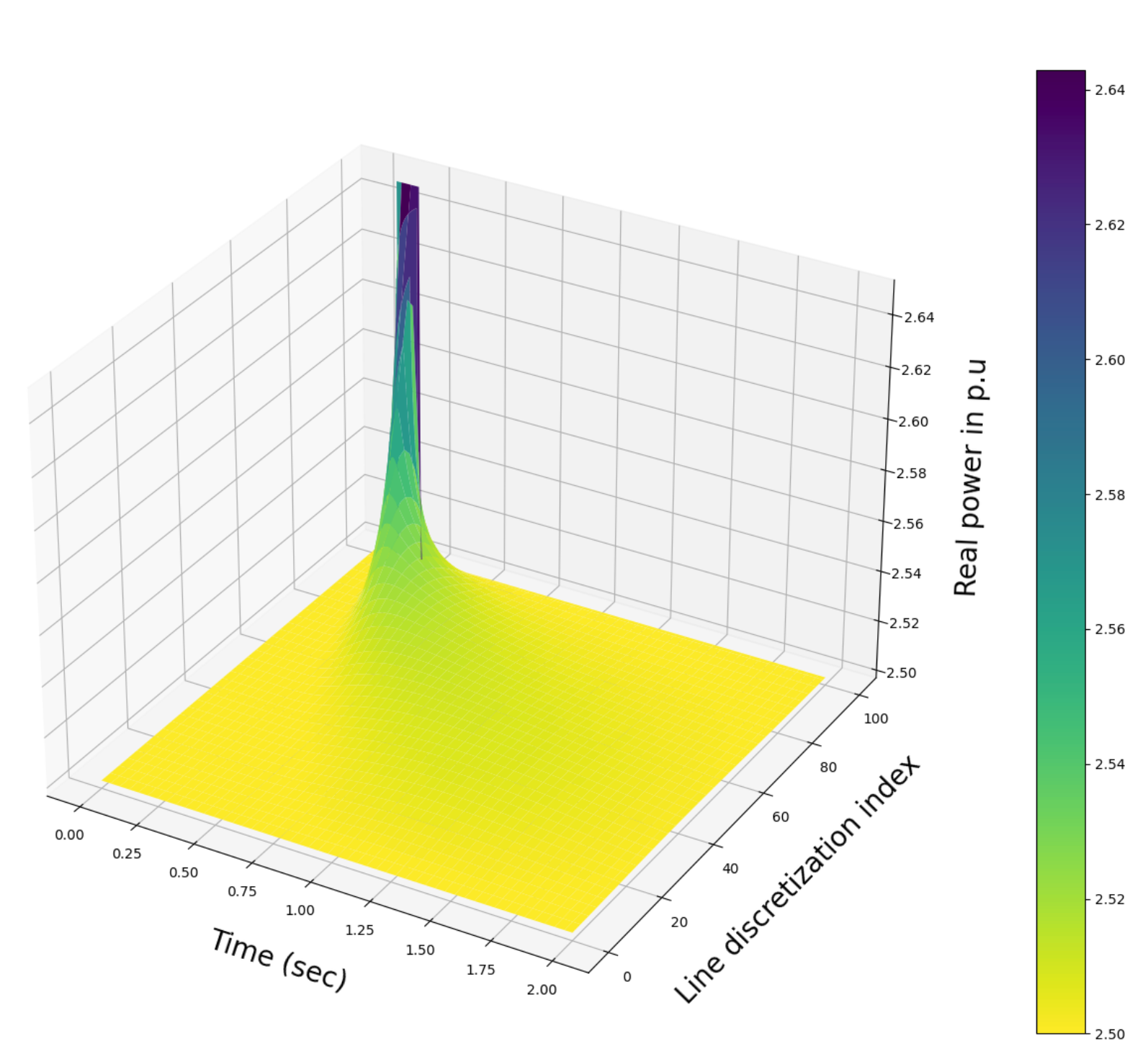}
    \caption{Electrical power response (H = 15 MJ/MVA.)}
  \end{subfigure}

  \caption{The electrical power response of the 2-bus power system to an increase in the active load power by 10\%.}
  \label{fig:myfig}
\end{figure}

To evaluate EMW propagation within this system, we first increase the active power load at Bus 1 by 10\%. Subsequently, we investigate the influence of the moment of inertia on the propagation of EMW by varying the inertia constant across the range of 1.5 to 15 MJ/MVA. Figures \ref{fig:2bussystemresults} and \ref{fig:myfig} display the responses of rotor angular velocity and electrical power, with the inertia moments of the generators set at 1.5 and 15, respectively. As observed, the system with lower inertia exhibits significantly larger deviations in both velocity and electrical power compared to the system with higher inertia. Additionally, the reduction in inertia amplifies the EMW oscillations. This underscores the potential for instability when a system has low inertia and experiences disturbances.

Therefore, these findings highlight the crucial importance of accurate EMW modeling, especially as renewable energy penetration increases. In such scenarios, the rapid response of protection systems becomes paramount in maintaining grid stability and preventing cascading failures as the system approaches critical conditions.

\subsection{The New England 39-Bus 10-Machine System}

\begin{figure}[htb!]
    \centering
    \rotatebox{0}{\includegraphics[scale=0.53, trim=2.5cm 16cm 2.5cm 2.5cm, clip]{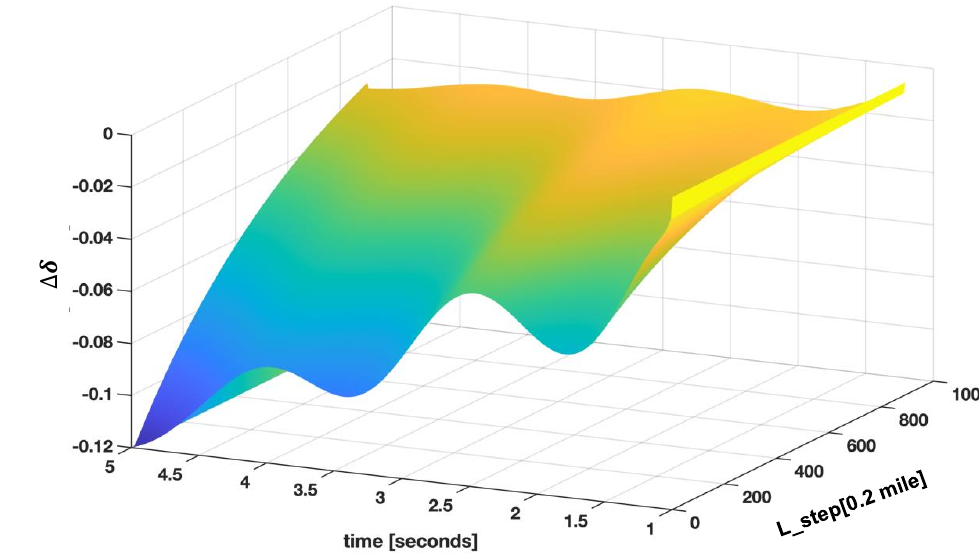}}
    \caption{EMW propagation from Bus 39 to Bus 31 using nonhomogeneous model.}
    \label{fig:3d}
\end{figure}

To evaluate the effectiveness of the proposed nonhomogeneous model method in a more realistic system, we conduct an EMW analysis on the 39-Bus power system. In this analysis, we introduced a 10\% increase in the load connected to Bus 39 as a disturbance. The process of determining the EMW propagation path from Generator 1 connected to Bus 39 and another generator follows a multistep procedure. Initially, we distribute the inertia characteristics of the generators. Subsequently, we use the inertia distribution method and the admittance values of the transmission lines to calculate the EMW velocity. Taking into account the physical length of the transmission lines and employing the Dijkstra algorithm, we determine the EMW propagation path starting from Bus 39. Table \ref{table:path2} shows the path of propagation of EMW from Generator 1 to Generator 2 connected to Bus 31.
\begin{table}[!hbt]
\centering
\caption{EMW propagation path from 39 to 31.}
\begin{tabular}{|c c c c c c|} 
 \hline 
Bus 39 $\rightarrow$&Bus 9 $\rightarrow$&Bus 8 $\rightarrow$&Bus 7 $\rightarrow$&Bus 6$\rightarrow$&Bus 31\\
  \hline  
\end{tabular}
\label{table:path2}
\end{table}\\
Simulations of EMW propagation from Bus 39 to Bus 31 are carried out employing the determined shortest path while taking into account boundary conditions and initial conditions. Figure \ref{fig:3d} displays a 3D representation of this EMW propagation using nonhomogeneous model. In this figure, the x-axis shows time in seconds and the y-axis represents the distance along the transmission line, with each step being 0.2 mile. It is important to note that when the disturbance is large enough, the EMW propagates through the transmission lines, reaching other synchronous generators and accelerating the rotor angle oscillation.

The magnitude and velocity of EMW propagation are not constant; they vary depending on the inertia and voltage characteristics of the transmission lines through which propagation occurs. Understanding how factors such as inertia and voltage influence the magnitude and velocity of EMW propagation is essential to evaluate the overall stability and reliability of the power system. This understanding enables a more comprehensive assessment of potential disturbances and their effects on the system's dynamic behavior.

\begin{figure}[htb!]
    \centering
    \rotatebox{270}{\includegraphics[scale=0.3, trim=2cm 1cm 1cm 0cm, clip]{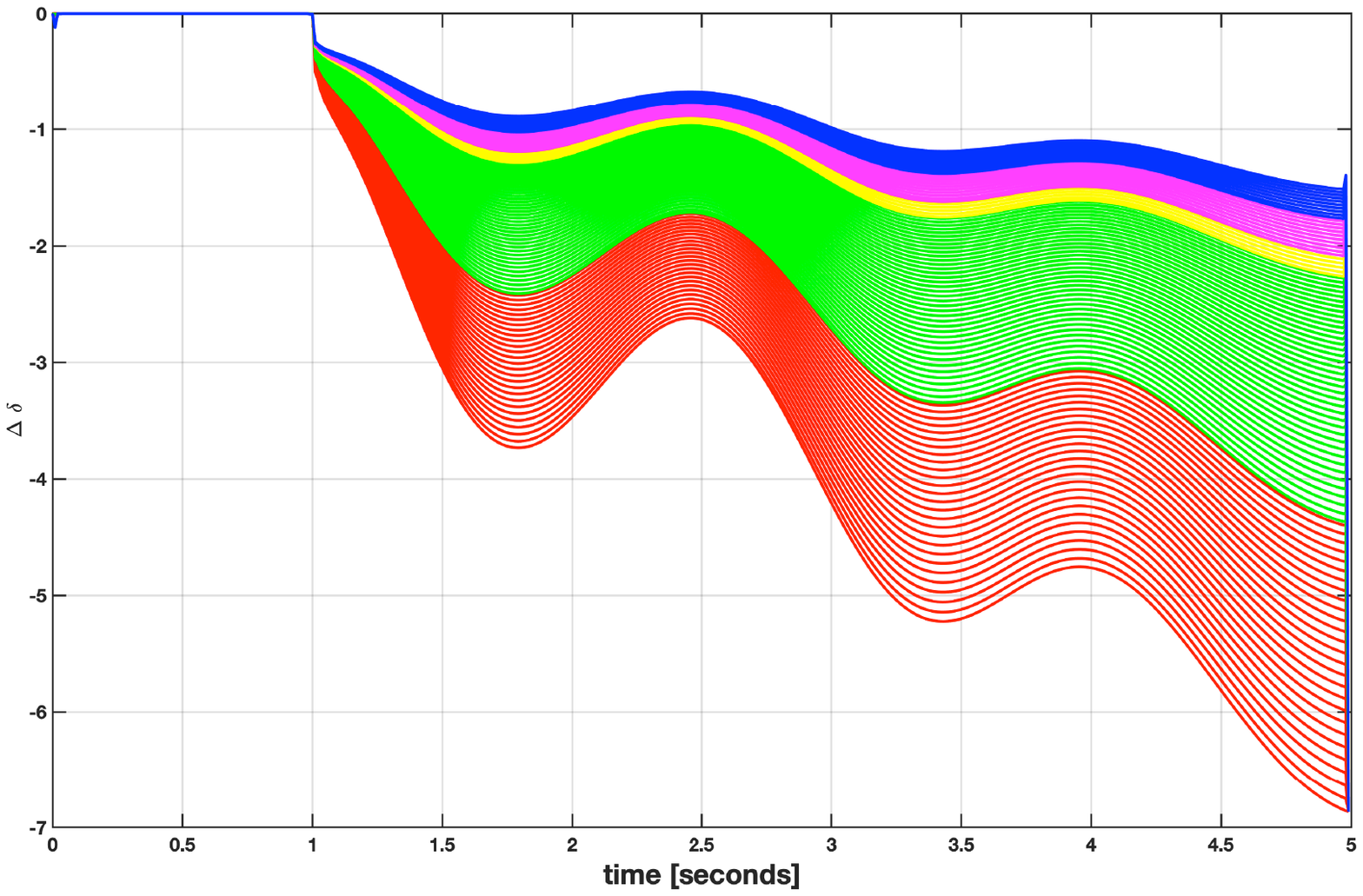}}
    \caption{ EMW propagation from Bus 39 to Bus 31 through lines with $\Delta \xi$ = 0.2 mile.}
    \label{fig:delta_delta}
\end{figure}

Figure \ref{fig:delta_delta} illustrates the EMW propagation at each step along the lines from Bus 39 to Bus 31 using the nonhomogeneous model. This two-dimensional representation corresponds to the 3D view in Figure \ref{fig:3d}. Each color in the figure represents a different transmission line. It is noted that the magnitude of the EMW decreases as it progresses through the transmission lines, while the oscillation frequency remains constant. In addition, EMW propagation is more perceptible near the fault location on the line between Bus 39 and Bus 9, indicated by the red lines in Figure \ref{fig:delta_delta}. This is due to certain initial and boundary conditions on Bus 39.

\begin{figure}[htb!]
    \centering
    \includegraphics[scale=0.07]{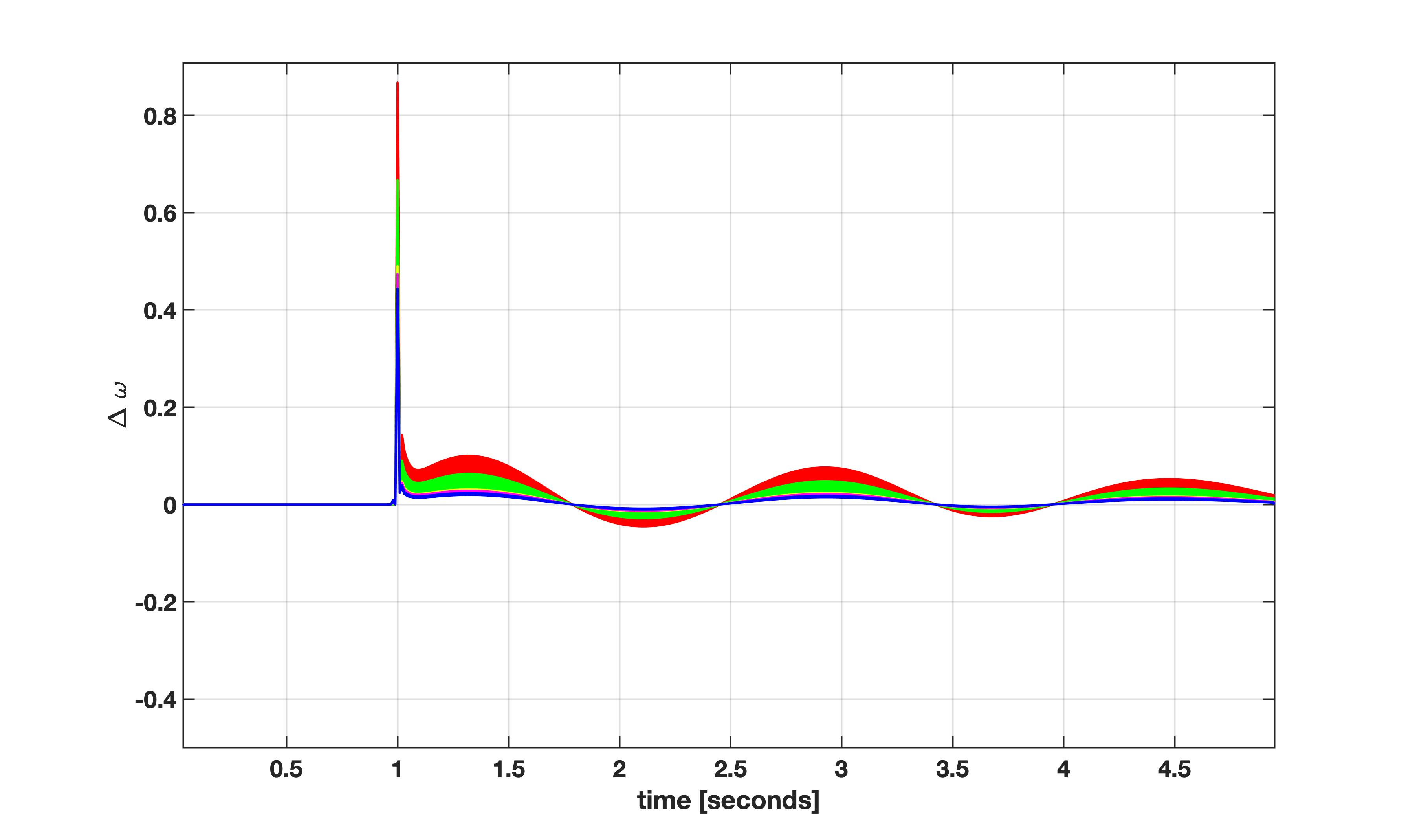}
    \caption{EMW propagation velocity from Bus 39 to Bus 31 after the disturbance.}
    \label{fig:velocity}
\end{figure}

Figure \ref{fig:velocity} illustrates the propagation characteristics of EMW in our study. As observed in the graph, the EMW propagation velocity exhibits a distinctive decline as they move away from the fault location to other points within the system. In particular, this reduction in velocity occurs while the EMW propagation frequency remains constant, maintaining a consistent frequency profile for the velocity. The observed damping in EMW oscillations during propagation through the transmission lines can be attributed to the inherent characteristics of the medium, which progressively attenuate the wave's energy as it travels.

Up to this point, we have shown the EMW propagation using the proposed nonhomogeneous model. In the next section, we will demonstrate the model's accuracy in depicting EMW propagation by comparing the results from homogeneous and nonhomogeneous models with PF simulation results. 

\subsection{Comparison Between Homogeneous and Non-homogeneous EMW Models}

\begin{figure}[htb!]

    \includegraphics[scale=0.08, trim=6cm 1cm 1cm 0cm, clip]{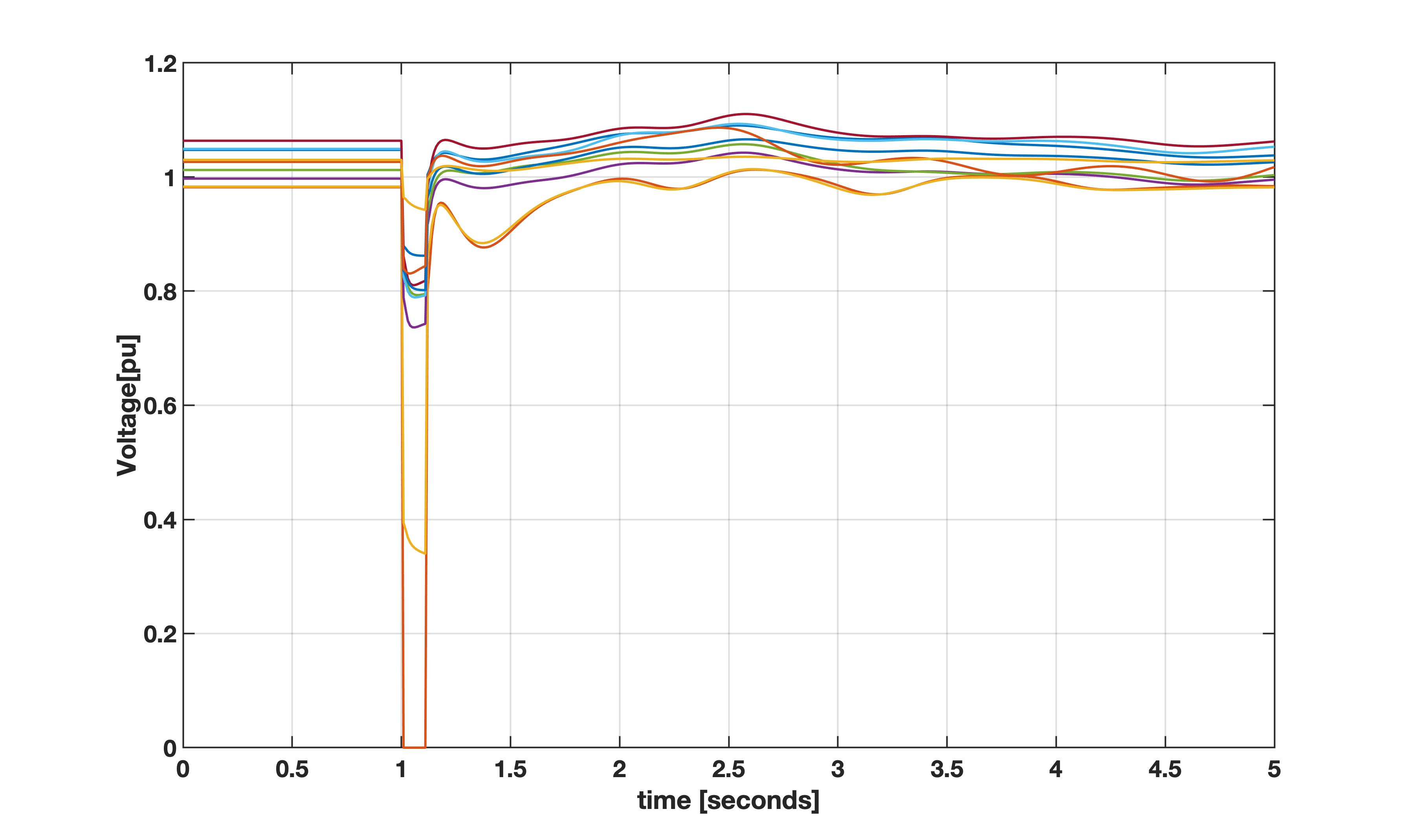}
    \caption{The variation of bus voltages during a 100 ms line outage disturbance occurring at time t=1s.}
    \label{fig:voltages}
\end{figure}

A significant difference arises between homogeneous and nonhomogeneous models, while the voltage of the power system drops due to faults such as large load shedding, generator outages, or short circuits. This finding has important implications for the accurate prediction and analysis of EMW behavior in power systems during large disturbances. To evaluate the performance of the proposed approach, we apply a line outage at Line 6-7 of the 39-Bus power system at time t=1s, which lasts 100 ms. As Figure \ref{fig:voltages} shows, all bus voltages, particularly those in close proximity to Bus 31, undergo a drop. It is important to note that during such occurrences, the assumption of unitary voltage magnitudes cannot be upheld.

\begin{figure}[htb!]
    \includegraphics[scale=0.170]{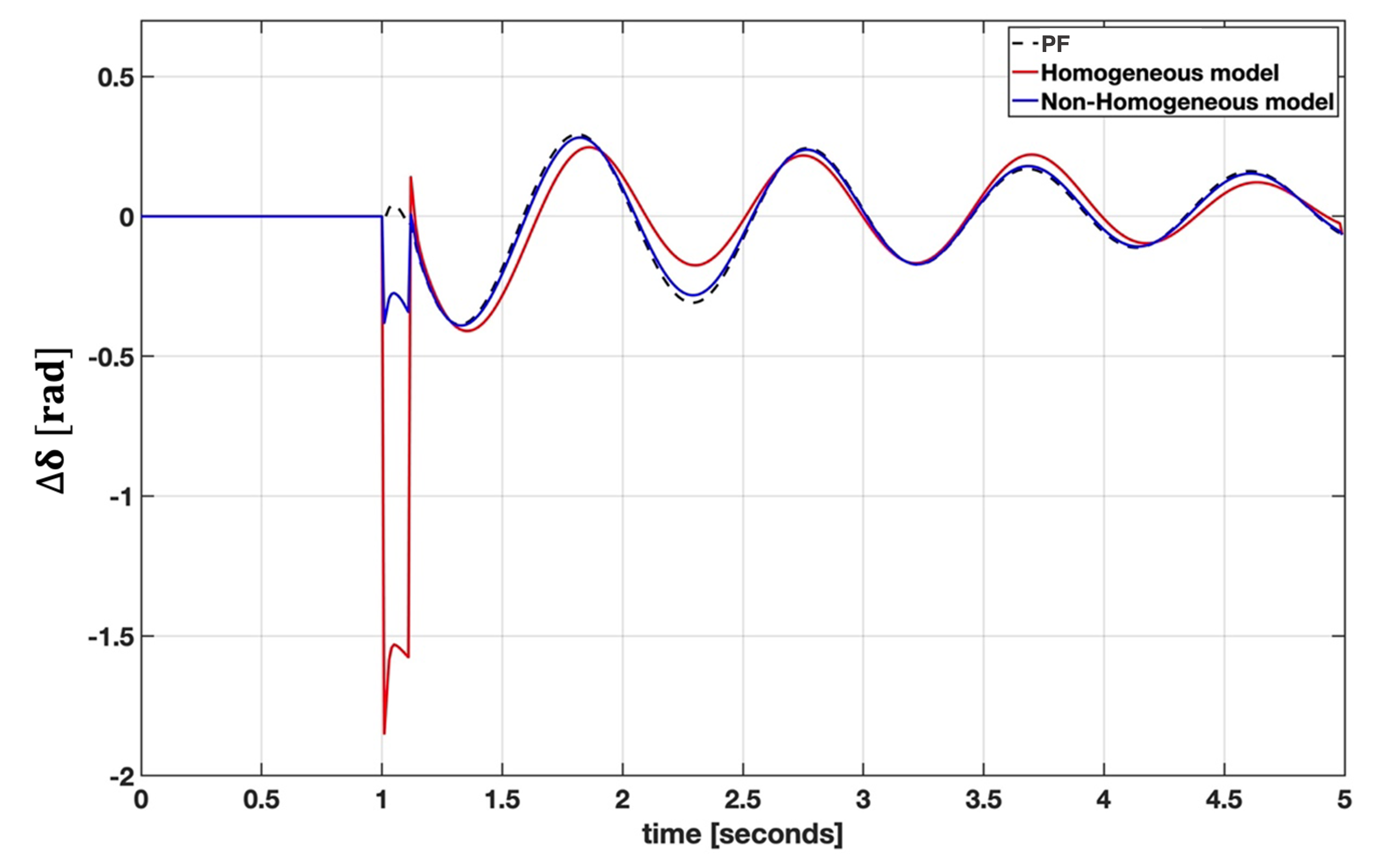}
    \caption{EMW oscillations at Bus 39 induced by a 100 ms line outage disturbance occurring at time t=1s.}
    \label{fig:rotor2}
\end{figure}

\begin{figure}[htb!]
    \includegraphics[scale=0.17]{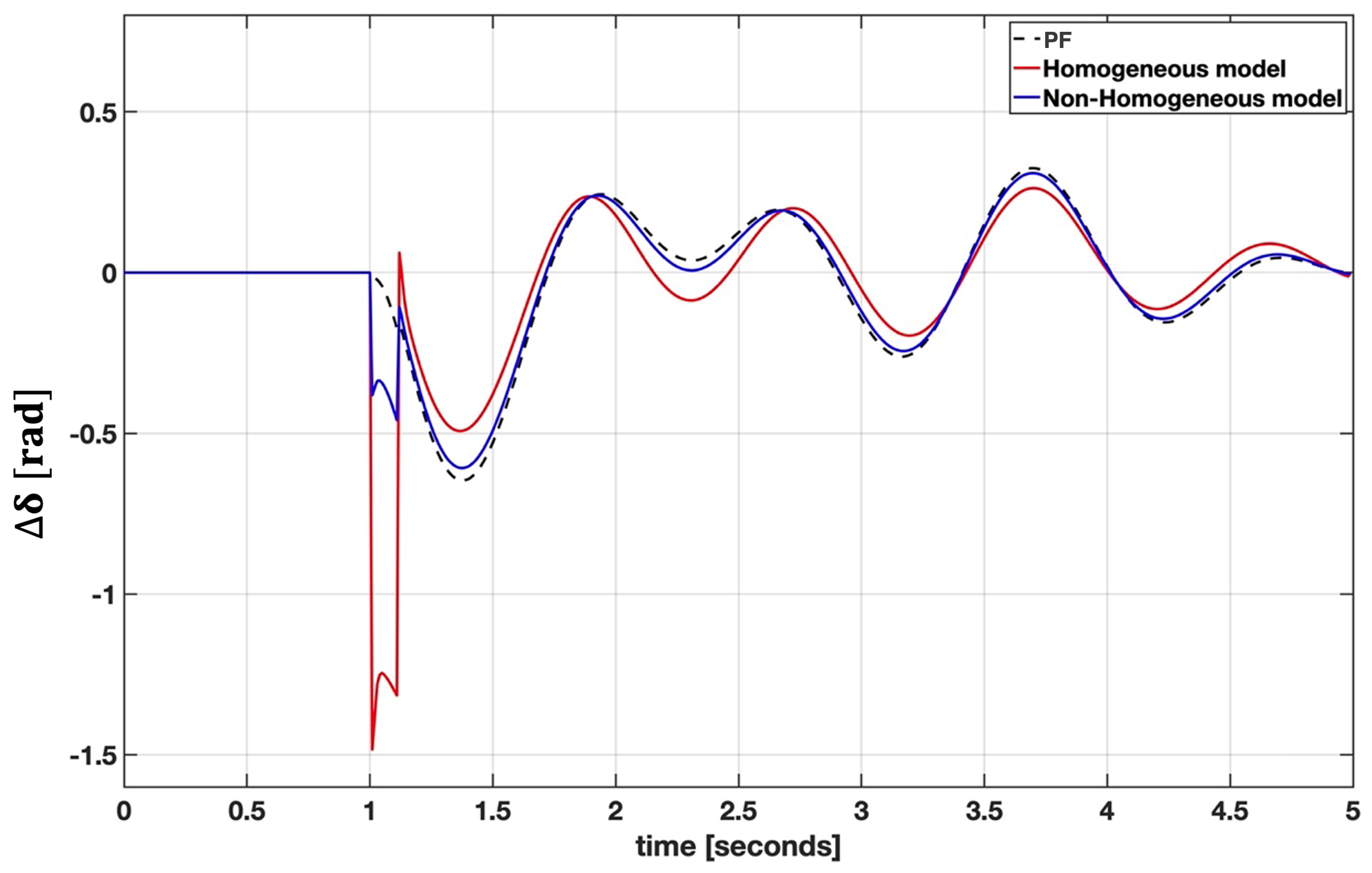}
    \caption{EMW oscillations at Bus 35 induced by a 100 ms line outage disturbance occurring at time t=1s.}
    \label{fig:velocity2}
\end{figure}

Figures \ref{fig:rotor2} and \ref{fig:velocity2} display the temporal variation of the EMW oscillations that originate from the generator on Bus 31 and to the generator connected to Bus 39 and Bus 37, respectively. They provide a comparison between the performance of homogeneous and nonhomogeneous EMW modeling approaches and the PF simulation results. In particular, the results highlight a substantial discrepancy between homogeneous and nonhomogeneous EMW modeling approaches compared to PF simulation results, underscoring the importance of considering voltage changes for more accurate representations of EMW propagation in power systems.

\begin{figure}[htb!]
    \includegraphics[scale=0.17]{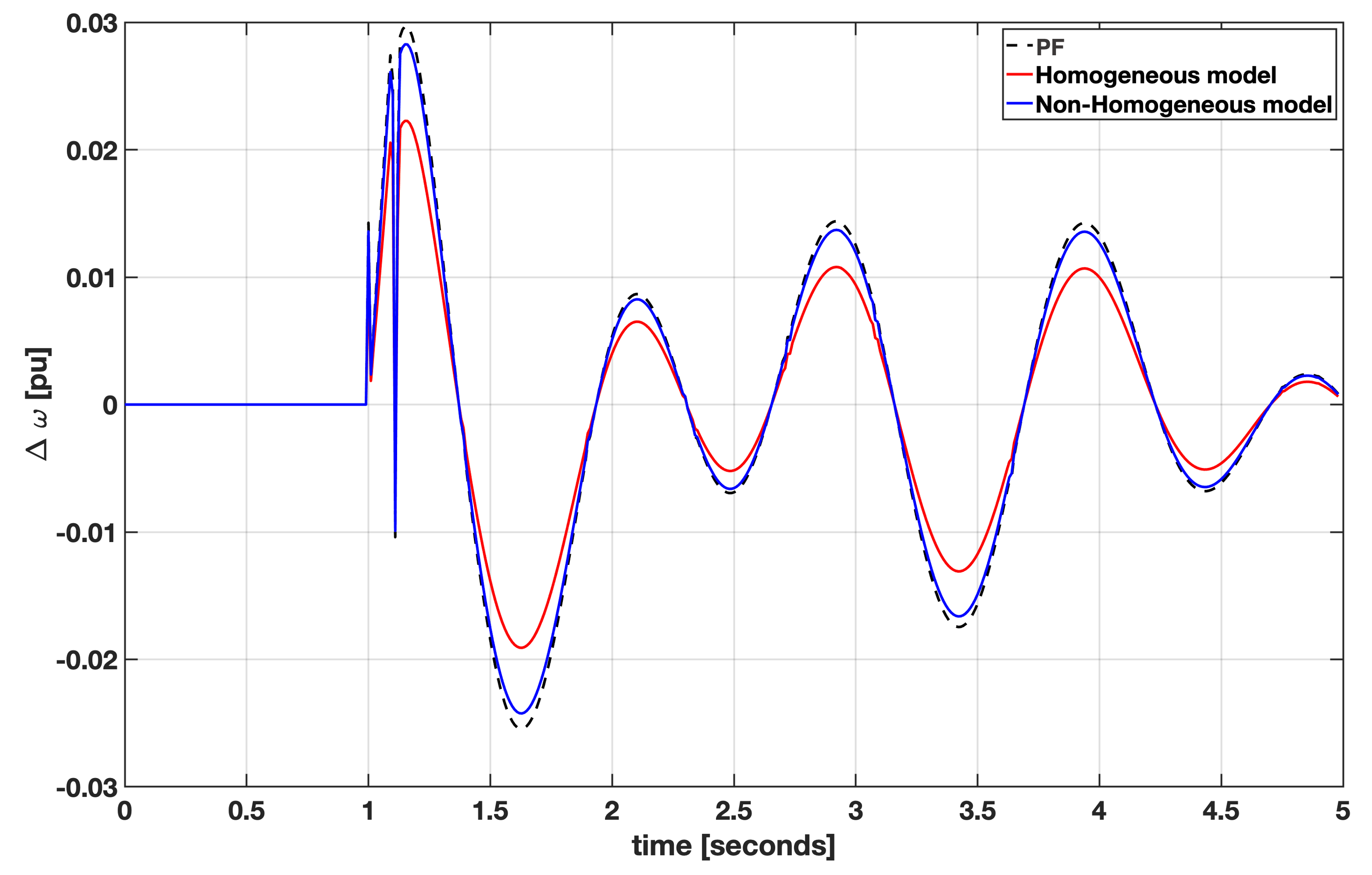}
    \caption{EMW velocity at Bus 39 induced by a 100 ms line outage disturbance occurring at time t=1s.}
    \label{fig:d1}
\end{figure}

\begin{figure}[htb!]
    \includegraphics[scale=0.17]{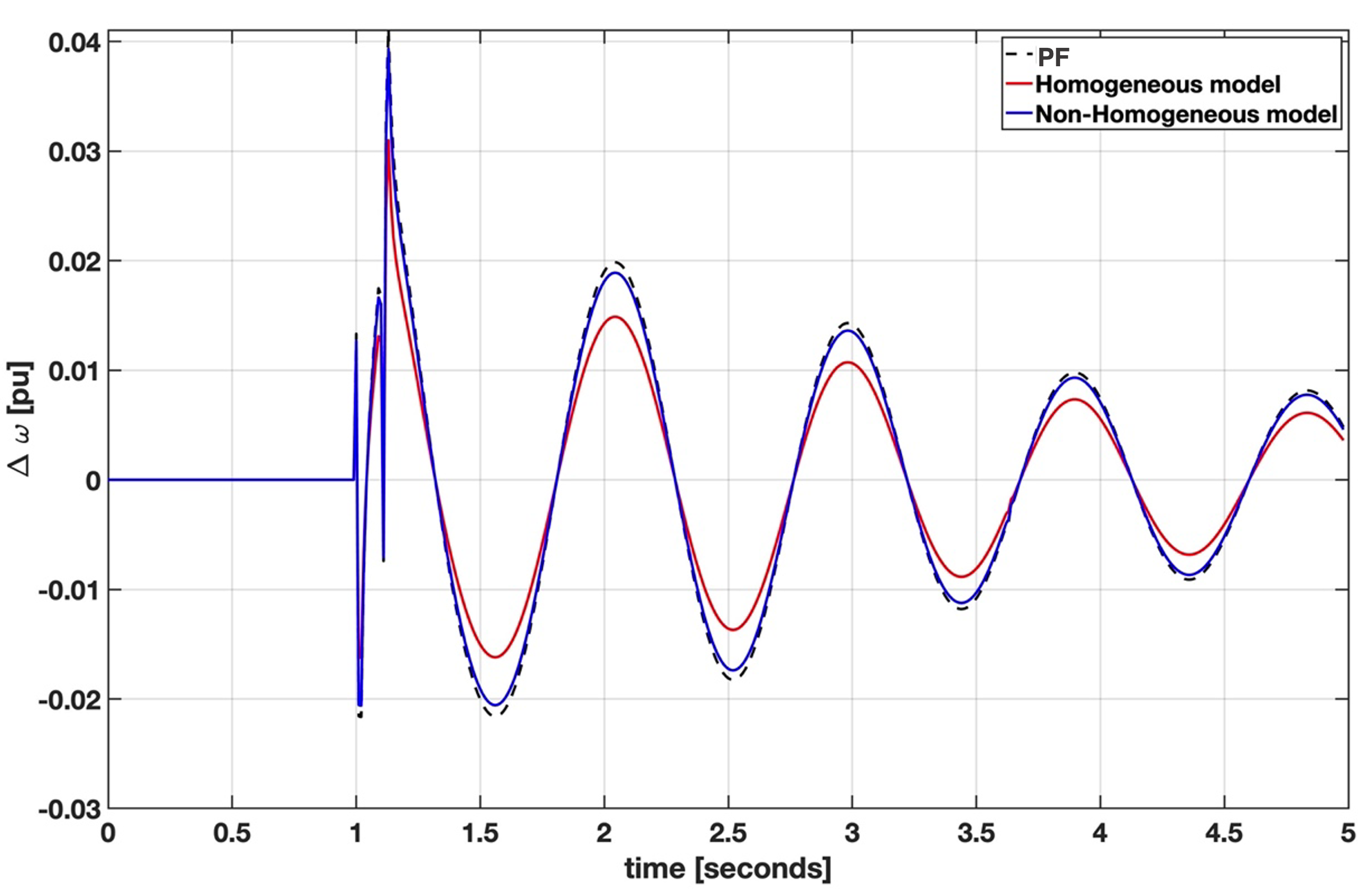}
    \caption{EMW velocity at Bus 35 induced by a 100 ms line outage disturbance occurring at time t=1s.}
    \label{fig:d2}
\end{figure}

To provide further clarity on this matter, a comparison between EMW velocities is also analyzed for this scenario. Figures \ref{fig:d1} and \ref{fig:d2}  show the difference between the EMW velocity at Bus 39 and at Bus 35 for the homogeneous and nonhomogeneous model and the PF simulation results. It is evident that the nonhomogeneous model not only offers a more accurate representation of EMW behaviors but also yields valuable insights into the system's susceptibility to voltage instability and transient events. These findings underscore the potential advantages of adopting a nonhomogeneous model to achieve greater precision and robustness in power system studies. Ultimately, this contributes to the enhancement of grid operation and overall reliability and resiliency.

\begin{figure}[htb!]
    \includegraphics[scale=0.075]{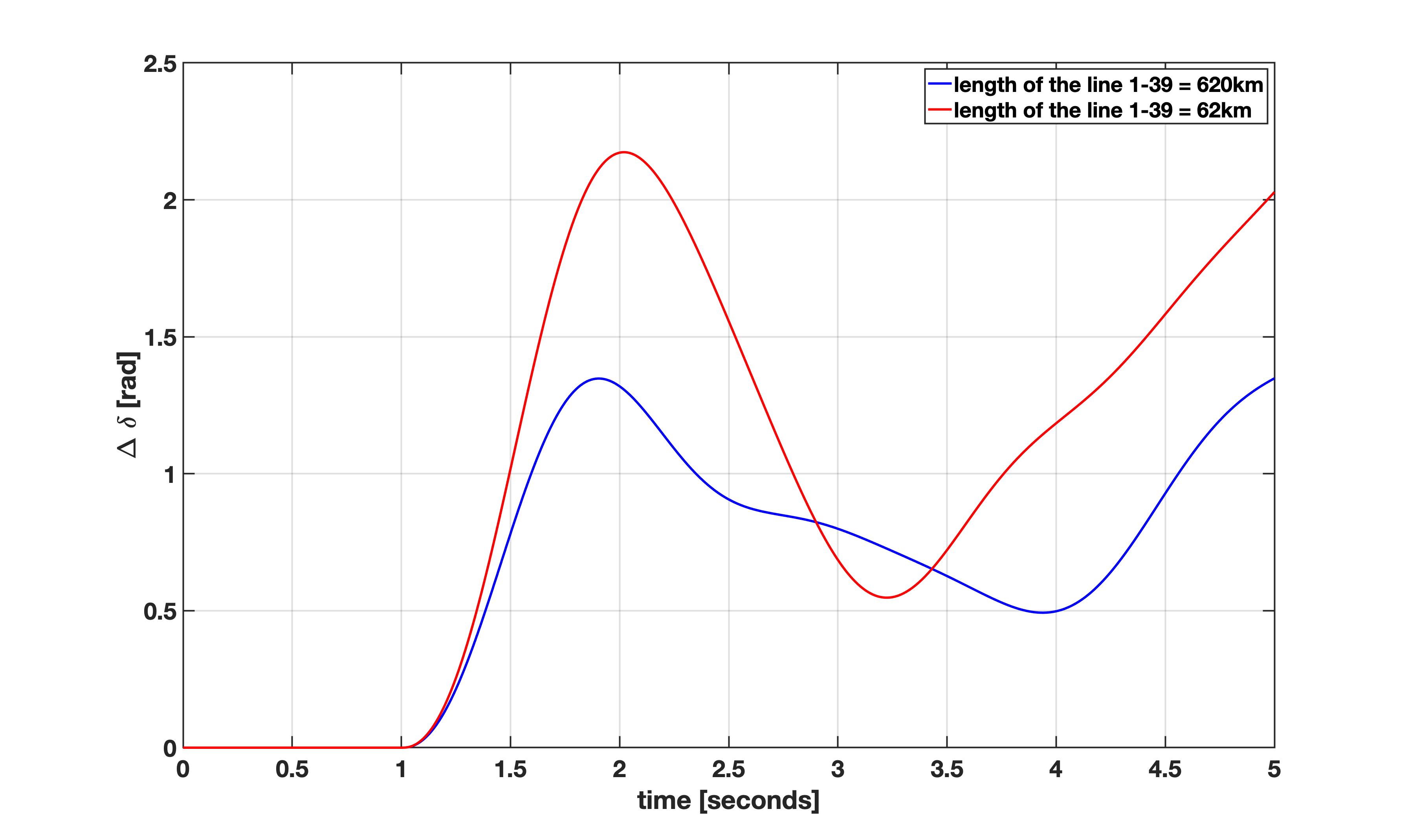}
    \caption{EMW at Bus 35 considering the load at Bus 39 increases by 10\% at time 1s induced by the changes of the length of the Line 8-9 from $l$(red) to $10l$ (blue).}
    \label{fig:d}
\end{figure}

To gain a deeper understanding of how EMWs behave concerning the length of transmission lines, we conduct a study examining the impact of line length on EMW propagation. To this end, we increase the length of the line connecting Bus 39 to Bus 1 by a factor of ten. Figure \ref{fig:d} illustrates the EMW propagation from Bus 39 to Bus 35 while subjecting the system to a 10\% increase in the load connected to Bus 39. As the figure shows, the length of the line affects EMW propagation. This result can be attributed to the fact that a line with a longer length exhibits greater associated inertia, as determined using the ABID method, making it more robust in the face of disturbances. Consequently, the EMW velocity decreases for lines with a larger length.

Comparison between the simulation results of the EMW modeling and the PF software has been performed only to show the validity of the EMW models. But we may say that EMW nonhomogenous modeling is more effective for analyzing the disturbance propagation, because using it we can predict disturbance arrival time and disturbance propagation velocity, and hence tune properly the relays and design resilient control systems to mitigate the risk of cascading failures that may lead to large-scale blackouts.

\section{Conclusion}

The modeling and analysis of EMW propagation in power systems are crucial to ensuring system stability and reliability, especially when there is a high penetration of renewable energy resources and power electronic devices. In this paper, a voltage-dependent model has been developed using a continuum modeling approach. The EMW partial differential equations have been solved using the Lax-Wendroff integration method. The impact of generator inertias has been investigated considering different inertia values. It has been shown that they increase not only the EMW propagation velocity but also the magnitude of the waves. Furthermore, our study includes a comparative analysis between our proposed model and a homogeneous model, which has revealed the limitations of the homogeneous model in accurately characterizing EMW behavior in the presence of voltage fluctuations. The impact of the lengths of the lines on the EMW has also been investigated. It has been concluded that the length of the line affects the EMW propagation when the ABID is used to distribute a moment of inertia in the continuum modeling approach.

In future work, we plan to compare the simulation results provided by the EMW non-homogeneous modeling and the PF software to the real data provided by PMUs deployed on actual power systems subjected to disturbances.

\color{black}

 



 







\end{document}